\documentclass[english,twocolumn,prb,aps,showpacs,floatfix, preprintnumbers,nofootinbib]{revtex4}
\usepackage{graphicx}%
\usepackage{dcolumn}
\usepackage{amsmath}
\usepackage{citesort}
\usepackage[dvips]{epsfig}
\usepackage[perpage,symbol*]{footmisc}
\usepackage{color}
\usepackage{babel}
\usepackage{prettyref}
\usepackage{float}
\usepackage{amsmath}
\usepackage{graphicx}
\usepackage[unicode=true, pdfusetitle,
 bookmarks=true,bookmarksnumbered=false,bookmarksopen=false,
 breaklinks=true,pdfborder={0 0 0},backref=false,colorlinks=true]
 {hyperref}
\hypersetup{
 linkcolor=blue, citecolor=black, anchorcolor=black,urlcolor=red}

\definecolor{LYXADDED}{RGB}{100, 50, 50}
\definecolor{LYXDELETED}{RGB}{50, 100, 50}

\usepackage{appendix}

\makeatother
\begin{document}
\vspace{1cm}

\title{Optimal coherent control of CARS: signal enhancement and background
elimination}

\author{Fang Gao}

\affiliation{Institute of Intelligent Machines, Chinese Academy of Sciences, Hefei
230031, China}

\author{Feng Shuang}

\email{fshuang@iim.ac.cn}

\affiliation{Institute of Intelligent Machines, Chinese Academy of Sciences, Hefei
230031, China}

\author{Junhui Shi}

\affiliation{Department of Chemistry, Princeton University, Princeton, NJ 08544,
USA}

\author{Herschel Rabitz}

\affiliation{Department of Chemistry, Princeton University, Princeton, NJ 08544,
USA}

\author{Haifeng Wang}

\affiliation{Department of Physics, National University of Singapore, Singapore
117542, Singapore}

\author{Ji-Xin Cheng}

\affiliation{Weldon School of Biomedical Engineering and Department of Chemistry,
Purdue University, West Lafayette, Indiana, 47907}

\date{\today}
\begin{abstract}
The ability to enhance resonant signals and eliminate the non-resonant background is analyzed for Coherent Anti-Stokes Raman Scattering (CARS).  The analysis is done at a specific frequency as well as for broadband excitation using femtosecond pulse-shaping techniques.  An appropriate objective functional is employed to balance resonant signal enhancement against non-resonant background suppression.  Optimal enhancement of the signal and minimization of the background can be achieved by shaping the probe pulse alone while keeping the pump and Stokes pulses unshaped.  In some cases analytical forms for the probe pulse can be found, and numerical simulations are carried out for other circumstances. It is found that a good approximate optimal solution for resonant signal enhancement in two-pulse CARS is a superposition of linear and arctangent type phases for the pump.  The well-known probe delay method is shown to be a quasi-optimal scheme for broadband background suppression. The results should provide a basis to improve the performance of CARS spectroscopy and microscopy.
\end{abstract}
\maketitle

\section{Introduction}

Coherent anti-Stokes Raman scattering(CARS), as a four-wave nonlinear
process\cite{Maker:1966p32736}, has been widely used in the past
few decades to study chemical systems in solutions, reactions in the gas
phase, and vibrational dynamics in gas and condensed phases.
CARS microscopy is a recently implemented technique for imaging biological
species, which was pioneered by Duncan \emph{et al}. using two dye
lasers and developed\cite{Duncan:1982p24639} by Xie \emph{et al}.
for high-sensitivity applications \cite{Cheng:2001p15976,Volkmer:2001p26451}.
As a combination of ultrafast nonlinear spectroscopy and microscopy,
CARS microscopy is a highly chemically selective and sensitive
technique that employs a CARS signal of an unlabeled sample and provides
higher spatial resolution than two-photon fluorescence microscopy.

When imaging biological samples, the typical width of Raman transitions
is a few wave numbers, so picosecond pulses are widely used in CARS
spectroscopy and microscopy\cite{Cheng:2001p15976,Cheng:2004p12944,Evans:2008p25862,leone2}.
Not only is the narrow bandwidth of the picosecond pulse adequate
to detect specific Raman bands, it also produces
low nonresonant background. On the other hand, when investigating
broadband CARS spectra\cite{Dudovich:2002p12846,Dudovich:2003p14516},
especially in the Raman fingerprint region which spans from 800 to
1800 $\mathrm{cm}^{-1}$, it is necessary to bring in broadband femtosecond
pulses. Femtosecond CARS can  be employed not only for direct imaging but also as a tool to determine some microscopic and macroscopic parameters, such as molecular anharmonicity\cite{motzkus1} and temperature\cite{motzkus1,motzkus2}.
However, this situation creats a dilemma when the pulse bandwidth
is  $\sim$ 1000 $cm^{-1}$, as the nonresonant background becomes
significant to possibly submerge the resonant signal and the fine vibrational
structure of CARS\cite{Oron:2002p14526,Oron:2002p14485}. The large
nonresonant contribution affects the shape of CARS spectra and complicates
data analysis, which becomes an obstacle for femtosecond CARS. Hence the
study of signal enhancement and background suppression is important
for effective Raman selective excitation in CARS, in which a broadband
femtosecond pulse is employed to excite multiple Raman modes\cite{Oron:2002p14485,Konradi:2005p11031,Konradi:2006p11028,Konradi:2006p11970,Zhang:2007p10884}.
Resonant signal enhancement and nonresonant background suppression
of CARS have been studied for many years. Polarization
CARS adjusts the polarization of the pump and Stokes pulses to suppress
nonresonant background\cite{Akhmanov:1977p24242,Oudar:1979p24306,Purucker:1993p24332,Cheng:2001p32647}.
Time-resolved CARS\cite{Materny:2000p24439} uses temporally
overlapped pump and Stokes pulses along with a delayed probe pulse to generate
a signal. This procedure  can also eliminate the nonresonant background, which is essentially instantaneous
while the resonant signal has a much longer decay time. Scully \textit{et
al.} proposed hybrid CARS\cite{Pestov:2007p23235}, in which the broadband
pump and Stokes pulses produce maximal Raman coherence and the
narrow-band time-delayed probe pulse suppresses the nonresonant background.
Cheng \emph{et al.} reported that epidetected CARS microscopy can
significantly reduce the solvent background\cite{Cheng:2001p15976}. The advent of spatial light modulators(SLMs)
has enabled coherent control and mode-selective excitation of CARS with femtosecond
pulses\cite{Wefers:1995p27478,Kawashima:1995p14548,Weiner:2000p25933,motzkus4}.
Background suppression without loss of the resonant signal has also been explored widely with the development
of femtosecond phase shaping techniques. By using SLM phase modulation and a variable wave plate,
Silberberg \emph{et al.} combined phase and polarization control
to yield background-free single-pulse multiplex CARS spectra with
high spectral resolution\cite{Oron:2003p14511}.
Leone \emph{et al.} combined interferometric and polarization/phase control to
demonstrate a method of single pulse interferometric CARS spectroscopy\cite{leone1,leone3},
which could extract the imaginary and real parts of the background-free resonant CARS spectrum in a single experimental measurement. Offerhaus \emph{et al.} employed vibrational phase contrast CARS, in which the measured phase components in the focal volume allows enhanced sensitivity and increased selectivity\cite{offerhaus1,offerhaus5}.
With the optical fields driving the CARS process and the local oscillator used for heterodyning both derived from a single beam by pulse shaping, Motzkus \textit{et al.} proposed highly sensitive single-beam heterodyne CARS microspectroscopy\cite{motzkus3}, in which the sensitivity of chemically selective detection at microscopic resolution is dramatically increased.
 It has also to be mentioned here that while phase shaping is effective to suppress nonresonant background in broadband CARS,
there are also post-processing approaches proposed by Bonn~\cite{bonn1,bonn2,bonn3}
and Cicerone\cite{cicerone1,cicerone2} to retrieve background-free and noise-reduced CARS spectra.

In this work, we explore the optimal control of the signal and background
of CARS with various phase shaping schemes. The paper is organized
as follows: Section \ref{sec:specific} shows the optimal phase shaping
schemes at a specific frequency. Section \ref{sec:Broad} investigates
the control strategies for broadband CARS and the conclusions are
given in Section \ref{sec:Conclusions}.

\section{Local optimal control\label{sec:specific}}

%
\begin{figure}[h]
\begin{centering}
\includegraphics[scale=0.3]{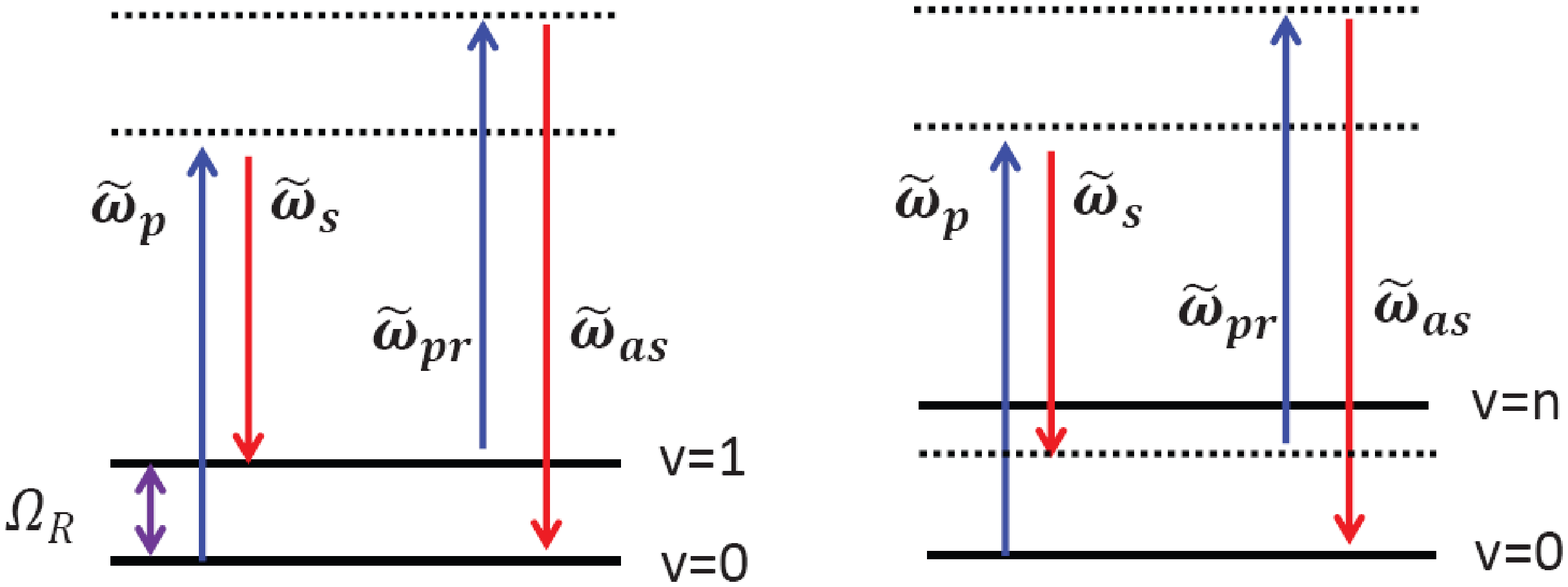}
\par\end{centering}

\caption{\label{fig:CARS-process}Energy level diagram of the CARS process.
The left panel corresponds to the resonant signal generation: pump and Stokes pulses generate coherence
between two vibrational levels, when they have a frequency difference
which coincides with the Raman resonance $\Omega_{R}$. The probe pulse then induces the anti-Stokes signal. The right panel corresponds to
the nonresonant background contribution: the nonresonant background is produced via
an intermediate virtual state that does not reflect the resonant
molecular energy level.}

\end{figure}

CARS is a four-wave nonlinear process as shown in Fig.\eqref{fig:CARS-process}.
Three laser pulses are used to produce the CARS signal: the pump pulse
$E_{p}(\omega_{p})$, the Stokes pulse $E_{s}(\omega_{s})$, and the
probe pulse $E_{pr}(\omega_{pr})$. The CARS signal $I_{\text{CARS}}(\tilde{\omega}_{as})$ is a coherent
superposition of resonant third order nonlinear polarization $P_{r}^{(3)}(\omega_{as})$
and nonresonant third order nonlinear polarization $P_{nr}^{(3)}(\omega_{as})$,

\begin{align}
{\scriptstyle I_{\text{CARS}}  (\tilde{\omega}_{as})=|P_{r}^{(3)}(\tilde{\omega}_{as})+P_{nr}^{(3)}(\tilde{\omega}_{as})|^{2}}\end{align}
and
\begin{equation}
\begin{split}
{\scriptstyle P_{\text{r}}^{\left(3\right)}\left(\tilde{\omega}_{as}\right)  =}& {\scriptstyle \iiint_{-\infty}^{+\infty}d\tilde{\omega}_{p}d\tilde{\omega}_{s}d\tilde{\omega}_{pr}{\textstyle \frac{C}{\Omega_{R}-\left(\tilde{\omega}_{p}-\tilde{\omega}_{s}\right)-i\Gamma}}}\\
&{\times \scriptstyle  \tilde{E}_{p}\left(\tilde{\omega}_{p}\right)
\tilde{E}_{s}^{*}\left(\tilde{\omega}_{s}\right)\tilde{E}_{pr}\left(\tilde{\omega}_{pr}\right)\delta(\tilde{\omega}_{as}-
\tilde{\omega}_{p}+\tilde{\omega}_{s}-\tilde{\omega}_{pr})}\label{eq:Pr}
\end{split}\end{equation}

\begin{equation}
\begin{split}
{\scriptstyle P_{\text{nr}}^{\left(3\right)}\left(\tilde{\omega}_{as}\right) =}& {\scriptstyle \iiint_{-\infty}^{+\infty}d\tilde{\omega}_{p}d\tilde{\omega}_{s}d\tilde{\omega}_{pr}\chi_{nr}}\\
&{\scriptstyle \times \tilde{E}_{p}\left(\tilde{\omega}_{p}\right)
\tilde{E}_{s}^{*}\left(\tilde{\omega}_{s}\right)\tilde{E}_{pr}\left(\tilde{\omega}_{pr}\right)\delta(\tilde{\omega}_{as}-\tilde{\omega}_{p}+\tilde{\omega}_{s}-\tilde{\omega}_{pr})\ }\label{eq:nonreosnant}
\end{split}\end{equation}
here $\Omega_{R}$ is the Raman frequency between energy level 1 and
2, $2\Gamma$ is the level width, $C$ is a constant which depends
on the material property, and $\chi_{nr}$ is the nonresonant third-order
susceptibility. The resonant signal intensity at the frequency $\tilde{\omega}_{as}$
is $|P_{r}^{(3)}(\tilde{\omega}_{as})|^{2}$, the nonresonant
background intensity is $|P_{nr}^{(3)}(\tilde{\omega}_{as})|^{2}$, and
the integrated CARS intensity is $I=\int_{0}^{+\infty}|P_{r}^{(3)}(\tilde{\omega}_{as})+P_{r}^{(3)}(\tilde{\omega}_{as})|^{2}d\tilde{\omega}_{as}$.

In our work, the carrier frequencies of the pump, Stokes and probe pulses
are denoted as $\Omega_{p}$, $\Omega_{s}$ and $\Omega_{pr}$, respectively.
Then if all the pulses are transform limited pulses(TLPs), the peak of the resonant and nonresonant
signals in the spectrum are both located at the frequency $\Omega_{as}=\Omega_{p}-\Omega_{s}+\Omega_{pr}$.

A Gaussian amplitude profile in the frequency domain is used for the pump,
Stokes and probe pulses in our theoretical treatment and simulations,

\begin{align}
{\scriptstyle \tilde{E_{k}}\left(\tilde{\omega}_{k}\right)  ={\textstyle \frac{E_{k}}{\Delta_{k}^{1/2}}}e^{-\left(\tilde{\omega}_{k}-\Omega_{k}\right)^{2}/\Delta_{i}^{2}}e^{i\tilde{\Phi}_{k}(\tilde{\omega}_{k}-\Omega_{k})}\text{,}\ k=\{P,\ S,\ Pr\},}\ \label{eq:GaussianPulse}\end{align}
where $2\sqrt{\ln2}\Delta_{k}$ and $2\sqrt{\ln2}\Delta_{k}$ are
the corresponding spectral full widths at half-maximum (FWHM), and
$\tilde{\Phi}_{k}(\tilde{\omega}_{k}-\Omega_{k})$ is the frequency-domain
phase profile. For simplicity, a frequency variable translation,
$\omega_{k}=\tilde{\omega}_{k}-\Omega_{k}$, is made (\emph{i.e.} $\Phi_{k}(\omega_{k})=\tilde{\Phi}_{k}(\tilde{\omega}_{k}-\Omega_{k})$
and $E_{k}(\omega_{k})=\tilde{E}_{k}\left(\tilde{\omega}_{k}\right)={\textstyle \frac{E_{k}}{\Delta_{k}^{1/2}}}e^{-\omega_{k}^{2}/\Delta_{k}^{2}}e^{i\Phi_{k}(\omega_{k})}$),
 and $P_{r}^{(3)}(\Omega_{as})$ is written
as $P_{r}$.

There are several experimental configurations for CARS: single-pulse
CARS\cite{Dudovich:2002p12846} (all three photons required are supplied
by the same short optical pulse), two-pulse CARS (the pump pulse also
acts as probe pulse in the experiment) and three-pulse CARS. In this
section, optimal control at a single signal frequency is investigated
in two-pulse and three-pulse CARS. Optimal control strategies with
the probe pulse are also explored. Only the
intensity at a specific frequency ($\tilde{\omega}_{as}=\Omega_{as})$
is considered throughout this section, and broadband optimal control
will be discussed in Section III.

For simplicity, unless otherwise stated, $\Delta_{p}=\Delta_{pr}=\Delta_{s}=\Delta$
is assumed in the following without loss of generality, as
the conclusions still hold when the bandwidths are different.

\subsection{Control of the signal or background by shaping the probe pulse in three-pulse
CARS\label{sub:Probe-only-Control-of}}

\begin{figure}[h]
\begin{centering}
\includegraphics[scale=0.3]{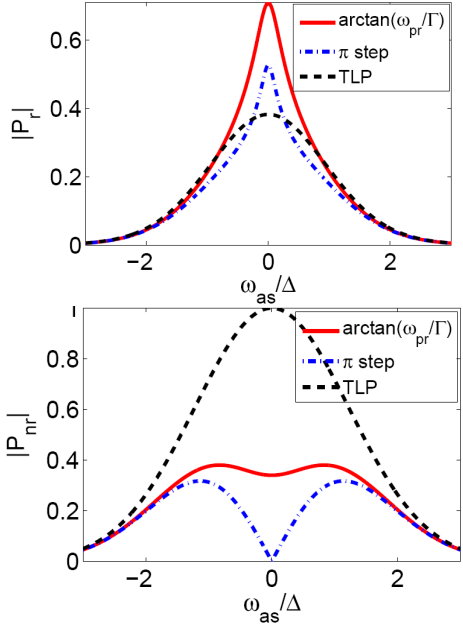}
\par\end{centering}
\caption{\label{fig:probe-only} The resonant signal and nonresonant background with different phase shaping schemes for the probe
pulse while keeping the pump and Stokes pulses unshaped in three-pulse CARS.
The arctan($\omega_{pr}/\Gamma$) phase (red solid lines)
, $\pi$-step phase (blue dash dotted lines) and TLP (black dashed lines) shaping schemes are shown together for comparison. The bandwidths of the pump, Stokes and probe pulses are the same: $\Delta_{p}=\Delta_{s}=\Delta_{pr}=\Delta=50cm^{-1}$.}
\end{figure}

Optimal control for signal enhancement or background suppression
with the probe pulse has been discussed previously\cite{arc1,arc2,arc3},
so only a brief description is given here. Fig.~\eqref{fig:probe-only}
shows the optimal control strategies: when the phase of the probe
pulse is either the arctan($\omega_{pr}/\Gamma$) or a $\pi$ step phase function,
the resonant and non-resonant signals achieve their maximal and minimal
values, respectively. The optimality condition can be gained analytically
as following:

Eq.\eqref{eq:Pr} leads to

\begin{align}
{\scriptstyle P_{r}} & {\scriptstyle =C\sqrt{\frac{\pi}{2\Delta}}\int_{-\infty}^{\infty}d\omega_{pr}\frac{1}{\omega_{pr}-i\Gamma}\exp
\left[-\frac{3\omega_{pr}^{2}}{2\Delta^{2}}\right]\exp\left[i\Phi_{pr}(\omega_{pr})\right]}\nonumber \\
 & {\scriptstyle =C\sqrt{\frac{\pi}{2\Delta}}\int_{-\infty}^{\infty}d\omega_{pr}\frac{1}{\sqrt{\omega_{pr}^{2}+\Gamma^{2}}}
 \exp\left[-\frac{3\omega_{pr}^{2}}{2\Delta^{2}}\right]\exp\left[i\left(\Phi_{pr}(\omega_{pr})+\alpha(\omega_{pr})\right)\right]}\label{eq:Probe-only}\end{align}
where $\alpha(\omega_{pr})=-\arctan(\omega_{pr}/\Gamma)+\pi/2$ is
confined in the domain $[0,\pi]$. It is easy to see that $|P_{r}|^{2}$
reaches its maximal value when the phase contribution to the integrand $\exp\left[i\Phi_{pr}(\omega_{pr})+i\alpha(\omega_{pr})\right]$
becomes a constant, \emph{i.e.} \begin{equation}
\Phi_{pr}(\omega_{pr})=\arctan(\omega_{pr}/\Gamma)+constant,\label{eq:arctan-phase}\end{equation}
with the resultant maximal peak intensity \begin{equation}
|P_{r}|_{max}^{2}=C^{2}\frac{\pi}{2\Delta}e^{\frac{3\Gamma^{2}}{2\Delta^{2}}}\mathrm{K}^{2}\left(0,\frac{3\Gamma^{2}}{4\Delta^{2}}\right),\label{eq:MaxPr-1}\end{equation}
Here $\mathrm{K}$ is a modified Bessel function. The condition for maximal or minimal resonant signal intensity could
also be established by using the variational method (see the appendix A),
which may aid in exploring the control landscape\cite{Rabitz:2004p1778}
for CARS.

The non-resonant background can be derived from Eq. \eqref{eq:nonreosnant}

\begin{align}
{\scriptstyle P_{nr}  (\Omega_{as})=\chi_{nr}\sqrt{\frac{\pi}{2\Delta}}\int_{-\infty}^{+\infty}d\omega_{pr}\exp
\left[-\frac{3\omega_{pr}^{2}}{2\Delta^{2}}\right]\exp\left[i\Phi_{pr}(\omega_{pr})\right]}\label{eq:-1}\end{align}
Thus $|P_{nr}(\Omega_{as})|^{2}$ reaches its minimal value of zero when
$\exp\left[i\Phi_{pr}(\omega_{pr})\right]$ is an anti-symmetric function,
\emph{e.g.} $\Phi_{pr}(\omega_{pr})$ is the $\pi$ step phase function
about $\omega_{pr}=0$ (note that $\exp\left[-\frac{3\omega_{pr}^{2}}{2\Delta^{2}}\right]$
is a symmetric positive-definite function).

As seen in the bottom panel of Fig.\eqref{fig:probe-only}, the $\pi$
step phase function can only eliminate the local component of background
around $\tilde{\omega}_{as}=\Omega_{as}$ to form a dip in the spectrum.
Thus the phase function obtained by minimizing
$|P_{nr}(\Omega_{as})|^{2}$ is \emph{locally} optimal, but the background
may still affect the resonant signal away from the position of $\tilde{\omega}_{as}=\Omega_{as}$
in the spectrum. Hence, a broadband background suppression method
is necessary for CARS, which will be discussed in Section III.

\subsection{Control in two-pulse CARS\label{sub:2pulseCARS}}

\begin{figure}[h]
\begin{centering}
\includegraphics[scale=0.3]{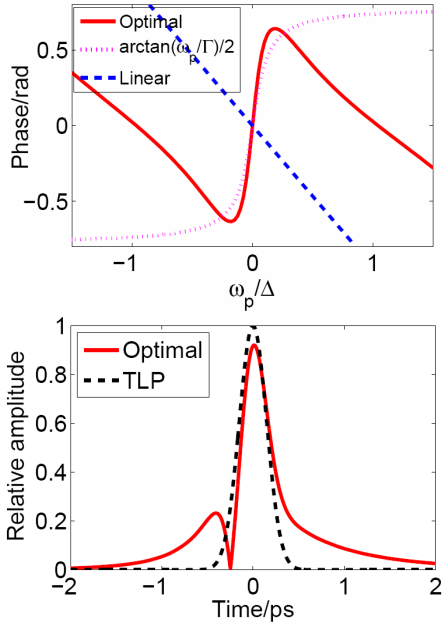}
\par\end{centering}

\caption{\label{fig:Two-pulse-CARS}Optimal phase function(via the CMA-ES optimization method\cite{Ostermeier:1994p25086,Hansen:2006p25132})
for the pump pulse in two-pulse CARS. The parameters: $\Delta_{p}=\Delta_{s}=50cm^{-1},$
$\Gamma=4.8cm^{-1}$. In the top panel, the red solid line is the
optimal phase function of the pump pulse for maximal resonant signal intensity,
the magenta dotted line corresponds to the arctan($\omega_{p}/\Gamma$)/2
phase, and the blue dashed line is a linear phase profile. The bottom
panel shows the outcome of optimal pulse (red solid line) and TLP (black dashed
line) in the time domain.}

\end{figure}

For practical considerations, in many CARS experiments, the pump pulse
also operates as the probe pulse. In this subsection, we will concentrate
on two-pulse CARS, in which only the pump pulse is phase shaped.
From Eq. \eqref{eq:Pr}, it follows that

\begin{equation}
\begin{split}
{\scriptstyle  P_{r}^{(3)}(\Omega_{as})} & {\scriptstyle =\int_{-\infty}^{\infty}\frac{C}{\omega_{pr}-i\Gamma}E_{p}\left(\omega_{pr}\right)
\left[\int_{-\infty}^{\infty}E_{p}\left(\omega_{p}\right)E_{s}^{*}
\left(\omega_{p}+\omega_{pr}\right)d\omega_{p}\right]d\omega_{pr}} \\
 & {\scriptstyle =\frac{1}{\Delta^{3/2}}\int_{-\infty}^{\infty}\frac{C}{\omega_{pr}-i\Gamma}e^{-\frac{3\omega_{pr}^{2}}{2\Delta^{2}}+i\Phi_{p}
 (\omega_{pr})}}\\
 &{\scriptstyle \times \left[\int_{-\infty}^{\infty}e^{-
 \frac{2(\omega_{p}+\omega_{pr}/2)^{2}}{\Delta^{2}}+i\Phi_{p}(\omega_{p})}d\omega_{p}\right]d\omega_{pr}}
 \end{split}\end{equation}
It is difficult to obtain an analytic optimal phase function for $|P_{r}^{2}(\Omega_{as})|$.
Instead, a numerical solution is presented in Fig.\eqref{fig:Two-pulse-CARS},
and it shows that the optimal phase function is approximately a superposition
of the linear and $\arctan(\omega_{p}/\Gamma)/2$ phases: the optimal
phase is quasi-linear away from $\omega_p=0$ and similar to $\arctan(\omega_{p}/\Gamma)/2$
around $\omega_p=0$. This behavior may be understood as follows: since the
zero phase profile (a special case of linear phase) is optimal for
$|\int_{-\infty}^{\infty}e^{-\frac{2(\omega_{p}+\omega_{pr}/2)^{2}}{\Delta^{2}}+i\Phi_{p}(\omega_{p})}d\omega_{p}|$
, and the $\arctan(\omega_{p}/\Gamma)$ phase is optimal for
$|\int_{-\infty}^{\infty}\frac{C}{\omega_{pr}-i\Gamma}e^{-\frac{3\omega_{pr}^{2}}{2\Delta^{2}}+i\Phi_{p}(\omega_{pr})}d\omega_{pr}|$
, then an approximate superposition of the linear and $\arctan(\omega_{p}/\Gamma)$
phase functions forms the optimal solution for $|P_{r}^{(3)}(\Omega_{as})|$,
which is the combination of these two integrals. With no explicit
analytical solutions, optimal control in two-pulse CARS has to be
studied numerically case by case. In an experiment, an approximate
superposition of a step and a proper time-delayed phase could effectively
produce a large resonant signal\cite{Oron:2002p14485} in the case of
small $\Gamma$.

\begin{figure}[h]
\begin{centering}
\includegraphics[scale=0.3]{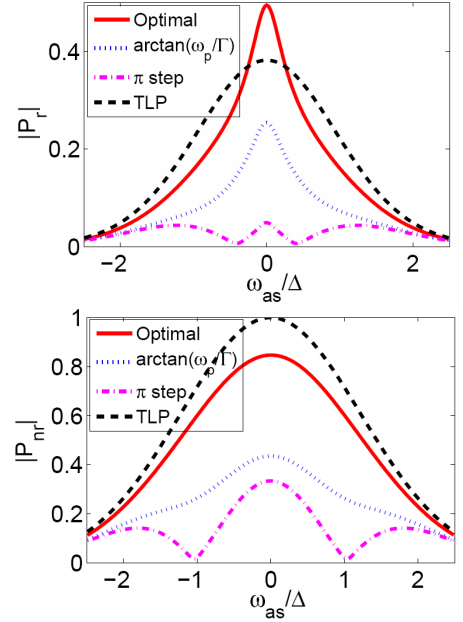}
\par\end{centering}

\caption{\label{Two-Pr-Pnr}
The resonant signal and nonresonant background spectra with
different phase shaping schemes for the pump pulse in two-pulse CARS. The optimal phase (red solid lines),
$\arctan(\omega_{p}/\Gamma)$ phase(blue dotted lines), $\pi$ step phase (magenta dash dotted lines) and TLP (black dashed lines)
schemes are shown together for comparison. The parameters
are the same as in Fig.~\ref{fig:Two-pulse-CARS}.
}
\end{figure}

To check the effect of this optimal phase scheme in two-pulse CARS, the resonant signal and nonresonant background
spectra with other shaping schemes are shown together in Fig.~\ref{Two-Pr-Pnr} for comparison. As can be seen, $|P_r|$ for the optimal scheme looks similar with that for the arctan($\omega_p/\Gamma$) scheme. The resonant signal decreases a lot with the $\pi$ step phase. Compared with the TLP scheme, all the other three schemes can suppress the background, but only the optimal phase scheme enhances the resonant signal. This is easy to understand since our optimized objective is the resonant peak signal at $\omega_{as}=0$.

\subsection{Control in three-pulse CARS\label{sub:3PulseCARS}}

In Sec. II(A), it was shown that the phase function $\arctan(\omega_{pr}/\Gamma)$
generates a maximal resonant signal intensity when only the probe pulse
is shaped. In three-pulse CARS, all the three pulses can be shaped.
Thus in this subsection, we will demonstrate if the three-pulse shaping
scheme can achieve better performance than the probe-only shaping
scheme in three-pulse CARS. The analytical and numerical results
will show that the configuration of TLPs for the pump and Stoke pulses
and an $\arctan(\omega_{pr}/\Gamma)$ phase profile for the probe pulse
is optimal to maximize the resonant signal. The CARS spectra of $|P_r|$ and $|P_{nr}|$
with this optimal phase scheme and other schemes can be found in Fig.~\ref{fig:probe-only}.

From Eq. \eqref{eq:Pr}, we have \begin{align}
{\scriptstyle P_{r}^{(3)}(\Omega_{as})} & {\scriptstyle =\int_{-\infty}^{\infty}\frac{C}{\omega_{pr}-i\Gamma}E_{pr}\left(\omega_{pr}\right)
\left[\int_{-\infty}^{\infty}E_{p}\left(\omega_{p}\right)E_{s}^{*}\left(\omega_{p}+\omega_{pr}\right)
d\omega_{p}\right]d\omega_{pr}}\nonumber \\
 & {\scriptstyle =\frac{1}{\Delta^{3/2}}\int_{-\infty}^{\infty}\frac{C}{\omega_{pr}-i\Gamma}e^{-
 \frac{3\omega_{pr}^{2}}{2\Delta^{2}}+i\Phi_{pr}(\omega_{pr})}A(\omega_{p},\omega_{pr})d\omega_{pr}}\label{eq:three}\end{align}
where the Raman excitation term is \begin{equation}
A(\omega_{p},\omega_{pr})=\int_{-\infty}^{\infty}e^{-\frac{2(\omega_{p}+\omega_{pr}/2)^{2}}{\Delta^{2}}}e^{i\Phi_{p}(\omega_{p})-i\Phi_{s}(\omega_{p}+\omega_{pr})}d\omega_{p}\end{equation}
It is easy to see that
\begin{equation}
\left|A(\omega_{p},\omega_{pr})\right|\leq\int_{-\infty}^{\infty}e^{-\frac{2(\omega_{p}+\omega_{pr}/2)^{2}}{\Delta^{2}}}d\omega_{p}=\sqrt{\frac{\pi}{2}}\Delta\end{equation}
The equality holds only when $\Phi_{p}(\omega_{p})-\Phi_{s}(\omega_{p}+\omega_{pr})$
does not depend on variable $\omega_{p}$ for arbitrary $\omega_{pr}$.
There are only two cases satisfying this condition: 1) $\Phi_{p}(\omega_{p})$
and $\Phi_{s}(\omega_{p}+\omega_{pr})$ are both constant functions\emph{,
i.e.} the pump and Stokes pulse are TLPs; 2) $\Phi_{p}$ and $\Phi_{s}$
are both linear functions with the same slope: $\Phi_{p}(\omega_{p})-\Phi_{s}(\omega_{p}+\omega_{pr})=k\omega_{pr}+constant$,
which is equivalent to linear phase shaping (or a time delay) scheme
for the probe pulse. The first case is just the second
one with zero slope (no delay).

\begin{figure}[h]
\begin{centering}
\includegraphics[scale=0.3]{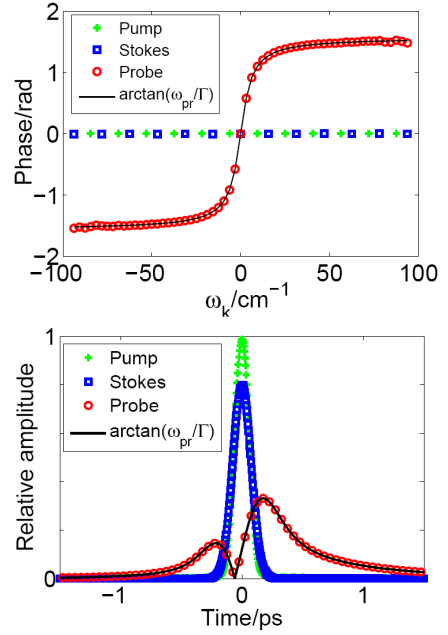}
\par\end{centering}

\caption{\label{fig:ThreePulse}Numerical optimal phase functions for achieving a maximal
resonant signal with three-pulse CARS using different pulse bandwidths:
$\Delta_{p}=125cm^{-1}$, $\Delta_{s}=100cm^{-1}$, $\Delta_{pr}=80cm^{-1}$.
The top and bottom panels correspond to the frequency and time domains,
respectively. The $\arctan(\omega_{pr}/\Gamma)$ phase (black
solid lines) is also shown for comparison.}

\end{figure}

Thus, the following equation can be derived
\begin{align}
 & \quad{\scriptstyle \left|\frac{1}{\Delta^{3/2}}\int_{-\infty}^{\infty}\tfrac{C}{\omega_{pr}-i\Gamma}
 e^{-\frac{3\omega_{pr}^{2}}{2\Delta^{2}}+i\Phi_{pr}(\omega_{pr})}A(\omega_{p},\omega_{pr})d\omega_{pr}\right|}\nonumber \\
 & {\scriptstyle  \leq\left|\frac{1}{\Delta^{3/2}}\int_{-\infty}^{\infty}|\tfrac{C}{\omega_{pr}-i\Gamma}
 e^{-\frac{3\omega_{pr}^{2}}{2\Delta^{2}}+i\Phi_{pr}(\omega_{pr})}|\left|A(\omega_{p},\omega_{pr})\right|d\omega_{pr}\right|}\nonumber \\
 & {\scriptstyle  \leq\left|\frac{1}{\Delta^{3/2}}\int_{-\infty}^{\infty}|\tfrac{C}{\omega_{pr}-i\Gamma}
 e^{-\frac{3\omega_{pr}^{2}}{2\Delta^{2}}+i\Phi_{pr}(\omega_{pr})}|\cdot\sqrt{\frac{\pi}{2}}\Delta\cdot d\omega_{pr}\right|}\nonumber \\
 & \quad (\mathrm{when}\ \Phi_{p}\ \mathrm{and}\ \Phi_{s}\ \mathrm{are\ constant})\nonumber \\
 & {\scriptstyle \leq C\sqrt{\frac{\pi}{2\Delta}}e^{\frac{3\Gamma^{2}}{4\Delta}}\mathrm{BesselK}\left(0,\frac{3\Gamma^{2}}{4\Delta^{2}}\right)}\nonumber \\
 & \quad  (\mathrm{when}\ \Phi_{pr}(\omega_{pr})=\arctan(\omega_{pr}/\Gamma))\end{align}
and the equality only holds when $\Phi_{p}$ and $\Phi_{s}$ are constant
and $\Phi_{pr}(\omega_{pr})=\arctan(\omega_{pr}/\Gamma)$. When $\Delta_{p}$,
$\Delta_{pr}$ and $\Delta_{s}$ are different, this conclusion still
holds. Hence the maximal resonant signal is only achieved when the
pump and Stoke pulse are unshaped TLPs, and the phase of the probe pulse
takes the form $\Phi_{pr}(\omega_{pr})=\arctan(\omega_{pr}/\Gamma)$.
This fact means that a two-pulse or three-pulse shaping scheme is
not necessary to achieve an optimal resonant signal, which was also verified
by numerical simulations as shown in Fig.\ref{fig:ThreePulse}. The
CMA-ES algorithm, which
is reliable for global optimization, is employed in the numerical
optimization. In agreement with the above analytical result, it is
found that the configuration of TLPs for the pump and Stoke pulses and
the $\arctan(\omega_{pr}/\Gamma)$ phase profile for the probe pulse is
optimal to maximize the resonant signal.

\subsection{Control of the signal-background difference by shaping the probe pulse\label{sub:Maximizing-the-difference}}

In the subsections above, the optimization of the resonant signal and
non-resonant background is treated separately. In the laboratory, however,
the signal and background are always detected together. As they can
not simultaneously reach extreme values, it is necessary to study
the balance between resonant signal enhancement and non-resonant
background suppression, which is a multi-objective optimization problem.
In this subsection, we will show how to achieve this goal by shaping
the probe pulse.

For this multi-objective problem, it is natural to consider the optimization
of the signal-to-background ratio. However, the ratio ($|P_{r}|/|P_{nr}|$)
is not a good choice: it could become infinite when $\left|P_{nr}\right|=0$,
no matter how small $\left|P_{r}\right|$ is. In this work, the difference of
the resonant signal and non-resonant
background intensities is chosen as the objective functional, \begin{equation}
J=\left|P_{r}\right|^{2}-k\left|P_{nr}\right|^{2}\label{eq:JOri}\end{equation}
where $k$ is the weight factor. This objective functional $J$ is
the balance of the minimization of $|P_{nr}|^{2}$ and maximization
of $|P_{r}|^{2}$. By maximizing the intensity difference, a large
signal-to-background ratio can be achieved with significant resonant signal intensity.

\begin{figure}[h]
\begin{centering}
\includegraphics[scale=0.3]{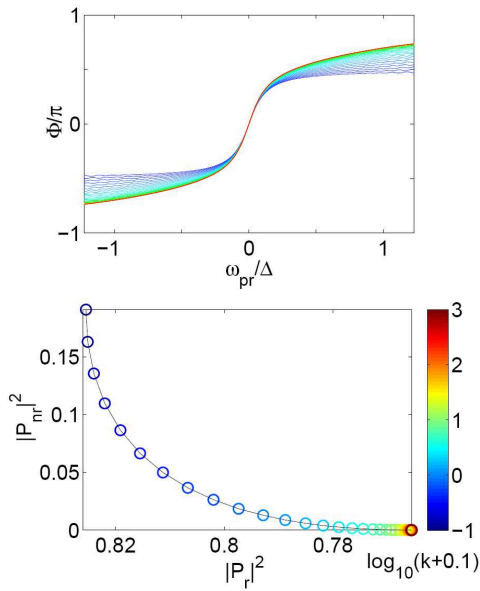}
\par\end{centering}

\caption{\label{fig:Optimal-different-k}The top panel: Numerical optimal
phase function of the probe pulse for $\left|P_{r}\right|^{2}-k\left|P_{nr}\right|^{2}$
with different weights $k$. The color of the lines indicates the value
of $k$, which is represented in the color bar on the right corresponding to $\log_{10}(k+0.1)$. All the phase
functions in this figure could significantly suppress the background. The bottom panel: The Pareto surface for the optimization
of signal enhancement and background suppression. With different weights
$k$, the value of $|P_{r}|^{2}$ is bounded in {[}0.765, 0.828{]},
while $|P_{nr}|^{2}$ is always much smaller than $|P_{r}|^{2}$. }
\end{figure}

According to the variational method, the necessary condition for a stationary
point of $J$ is
\begin{equation}
\frac{\delta J}{\delta\Phi_{pr}(\omega_{pr})}=0.\end{equation}
Numerical results indicate that the optimal phase $\Phi_{pr}(\omega_{pr})$
is an odd function of $\omega_{pr}$. After a detailed analysis
shown in appendix B, it is found that the optimal phase for maximizing
$J$ is a modified arctan-type function,

\begin{equation}
\Phi_{pr}(\omega_{pr})=\arctan\left(\frac{\omega_{pr}}{\Gamma-\lambda\gamma(\omega_{pr}^{2}+\Gamma^{2})}\right)+\theta,\label{eq:Pr-Pnr-phase}\end{equation}
where $\lambda=k\left(\chi_{nr}/C\right)^{2}$, $\omega_{pr}=\tilde{\omega}_{pr}-\Omega_{pr}$,
$\gamma$ (dependent on $\lambda$) is a parameter determined in Eq.
\eqref{eq:gamma} of appendix B, and $\theta$ is a trivial phase
angle. When the weight factor $k=0$, the optimization of $J$
reduces to the maximization of resonant signal, and its solution $\arctan(\omega_{pr}/\Gamma)$
is also consistent with the result in Eq. \eqref{eq:arctan-phase}.
With parameters $C=1$ and $\chi_{nr}=0.1$, the numerical optimal phases
for maximizing $\left|P_{r}\right|^{2}-k\left|P_{nr}\right|^{2}$ with
different $k$ are shown in the top panel of Fig.\eqref{fig:Optimal-different-k}.
It is easy to see that the Pareto surface of $J$ in the bottom panel
of Fig.\eqref{fig:Optimal-different-k}, which is defined as the set of optimal points ($|P_r(k)|$, $|P_{nr}(k)|^2$) for maximal $J$ with different $k$,  has two limit points. In the first limit, the pure maximization of $|P_{r}|^{2}$ without
considering $|P_{nr}|^{2}$ leads to $|P_{r}|^{2}$ reaching its maximal value
of 0.828 while $|P_{nr}|^{2}$ has a considerable value of 0.19, and
the phase function is $\arctan(\omega_{pr}/\Gamma)$. If $\chi_{nr}$
is large in this case, $|P_{nr}|^{2}$ is large as well, \emph{e.g.}
when $\chi_{nr}=1.0$, then $|P_{nr}|^{2}=1.9$. Hence the choice of $k=0$
will not achieve an optimal balance of resonant signal enhancement and
non-resonant background suppression in general cases with a large background. In the second case
when $k$ becomes large, then $|P_{r}|^{2}$ approaches its lower
limiting value of 0.765 while $|P_{nr}|^{2}$ is very small, and the
phase function also approaches a limiting curve as indicated with the red
color in the top panel of Fig.\eqref{fig:Optimal-different-k}. In
this case, the background could almost be eliminated while keeping
a considerable resonant signal intensity. Sec. II(A) showed the $\pi$
step phase could also eliminate the background, but it produced a relatively
small resonant signal intensity $|P_{r}|^{2}=0.45$(with the same
parameters as here). Hence, the $\pi$ step phase is not a good Pareto
optimal solution for the optimization of $J$, although it is widely
used in experiments.

From the analysis above, a practical choice for $k$ can be made:
1) when $\chi_{nr}/C$ is large (\emph{i.e.} a large background), it
is better to choose a large value of $k$, which will give a solution
close to the limit point on the right side of the Pareto surface;
2) when $\chi_{nr}/C$ is small, a small value of $k$ is appropriate.
For example, when $\chi_{nr}/C$ approaches zero, the $k$=0 is the best
choice, in which case there is no need to consider further reducing background.

\section{Broadband optimal control\label{sec:Broad}}

In Section II, resonant signal enhancement and non-resonant background
suppression at a specific frequency was studied. However, in many experiments,
a large resonant signal and low non-resonant background over the entire
spectrum is designed, so the local optimal control is not adequate.
In this section, we will show how to achieve broadband optimal
control. The $\pi$ step phase, which achieves perfect local
elimination of the background, will be investigated again, and a multi-$\pi$
step phase scheme is further proposed to yield better broadband background suppression. Finally, a global numerical
optimized phase profile for broadband background suppression will
be shown by employing a specific objective functional.

\subsection{Multi-$\pi$ step phase scheme}

\begin{figure}[h]
\includegraphics[scale=0.3]{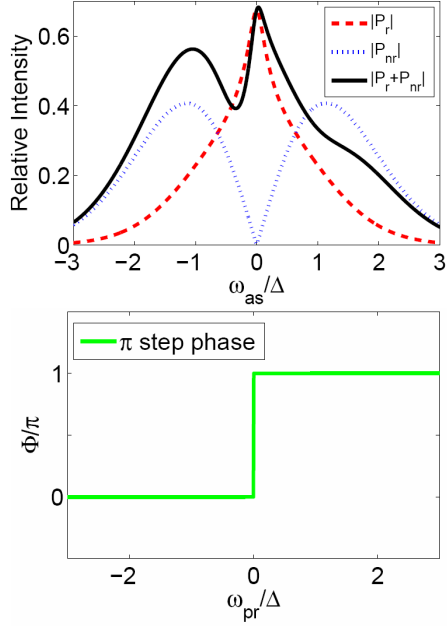}

\caption{\label{fig:single-pi-step-phase}The top panel: The resonant signal(red
dashed line), background(blue dotted line), and the whole CARS signal(black
solid line) with the $\pi$ step phase scheme; The bottom panel: The
$\pi$ step phase profile of the probe pulse which steps about $\omega_{pr}=0$.
Parameters: $\Delta=50cm^{-1}$ , $\Gamma=4.8cm^{-1}$. }

\end{figure}

It is shown above that the $\pi$ step phase($\mathrm{Heaviside}(\omega_{pr})\cdot\pi$)
could completely eliminate the background at $\Omega_{as}$,which
can also be achieved at another specific frequency by changing the $\pi$
phase step position. The effects of the $\pi$ step phase function
are shown in Fig.\eqref{fig:single-pi-step-phase}. As can be seen,
this phase scheme could not effectively eliminate the background at
other frequency components at the same time, although it produces
a good signal-to-background ratio at $\Omega_{as}$, which means that the
$\pi$ step phase is not a good choice for broadband background
elimination.

\begin{figure}[h]
\begin{centering}
\includegraphics[scale=0.3]{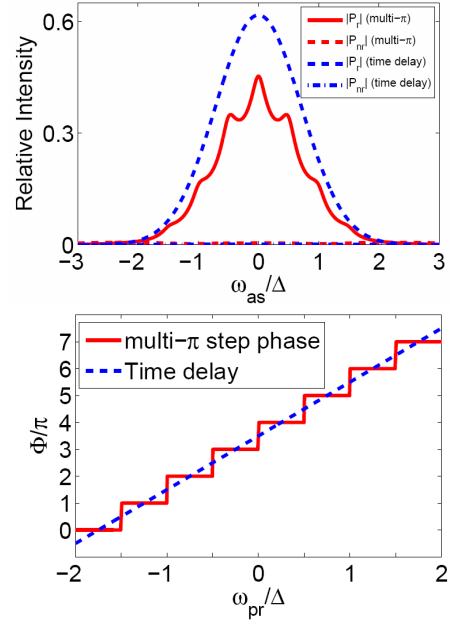}
\par\end{centering}

\caption{\label{fig:multi-pi-step}The top panel: The resonant signal intensity
$|P_{r}|$ and non-resonant background $|P_{nr}|$ intensity with
the multi-$\pi$ step phase scheme (red lines) and the time delay scheme
(blue lines). The bottom panel: The multi-$\pi$ step phase profile
(red solid line) and the time delay phase profile (blue dashed line).}

\end{figure}

Since a single $\pi$ step phase could eliminate the background locally
and its effective region for background elimination is only about
$\Delta/2$ as shown Fig.\eqref{fig:single-pi-step-phase}, it is
natural to construct a multi-$\pi$ step phase with jumps at several
positions for broadband background suppression. Every $\pi$ phase
step corresponds to a local region for background suppression. Fig.\eqref{fig:multi-pi-step}
illustrates a multi-$\pi$ step phase scheme which consists of eight
$\pi$ steps with an equivalent spacing of $\Delta/2$. With this
ladder-like multi-$\pi$ step phase, the background is perfectly suppressed
and the resonant signal remains large. The ladder shape of the multi-$\pi$
step phase is close to a time delay scheme as shown by the two curves
in the bottom panel of Fig.\eqref{fig:multi-pi-step}. In fact, similar CARS spectra
are yielded by these two schemes as shown in the top panel of Fig.\eqref{fig:multi-pi-step},
which indicates that the time delay scheme may be good for
broadband background elimination. This will be further demonstrated
in the following subsection by numerical global optimization.

\subsection{Numerical global optimization\label{sub:Numerical-global-optimization}}

\begin{figure}[h]
\begin{centering}
\includegraphics[scale=0.3]{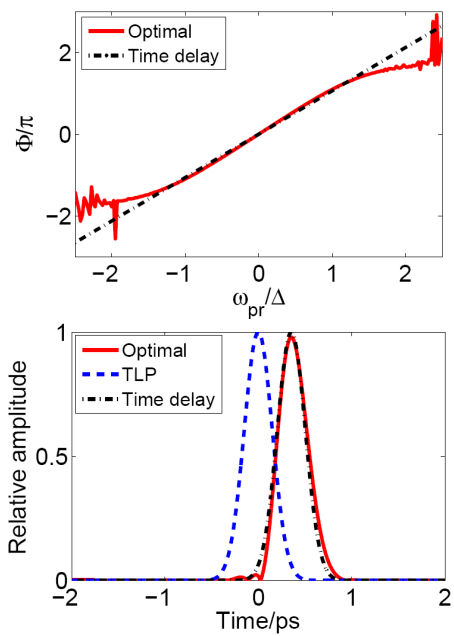}
\par\end{centering}

\caption{\label{fig:Optimal-phase-for-Ir-Inr}Optimal phase profile (with the
CMA-ES algorithm) for the maximization of $I_{r}-I_{nr}$. Because
the numerical optimal phase function for the pump and Stokes pulses are
zero phase, \emph{i.e.} TLP, they are not shown in this figure. The
top and bottom panels show the phase function and amplitude for the probe pulse(red
solid lines) in the frequency and time domain, respectively. The unshaped
TLP (blue dashed line) and quasi-optimal time delay (black dash dotted lines) schemes are also shown for comparison.
Parameters: $\Delta=50cm^{-1}$ and $\Gamma=4.8cm^{-1}$. }

\end{figure}

Since it is difficult to obtain an analytical optimal phase function
for broadband background suppression, a global numerical optimization
of the phase profile with a specific objective functional is performed,
which is defined as the difference of the integrated resonant and non-resonant signal intensities.

\begin{eqnarray}
J[\Phi_{pr},\Phi_{p},\Phi_{s}] & = & I_{r}-I_{nr},\label{eq:J-I-diff}\\
I_{r} & = & \int_{-\infty}^{+\infty}|P_{r}(\tilde{\omega}_{as})|^{2}d\omega_{as},\\
I_{nr} & = & \int_{-\infty}^{+\infty}|P_{nr}(\tilde{\omega}_{as})|^{2}d\omega_{as},\end{eqnarray}
 The optimization of this objective functional aims for a
large resonant signal while suppressing the background. Numerical optimization
with respect to all three pulses is carried out with $\Delta=50cm^{-1}$
and $\Gamma=4.8cm^{-1}$. The optimal phase shaping configuration
consists of unshaped pump and Stokes pulses and a shaped probe pulse
as shown in Fig.\eqref{fig:Optimal-phase-for-Ir-Inr}. It can be seen
that the optimal phase profile is quasi-linear.  In other words, the time delay scheme,
a well-known method for the background suppression, is quasi-optimal
for broadband background elimination with
the objective functional defined in Eq. \eqref{eq:J-I-diff}.

\begin{figure}[h]
\begin{centering}
\includegraphics[scale=0.3]{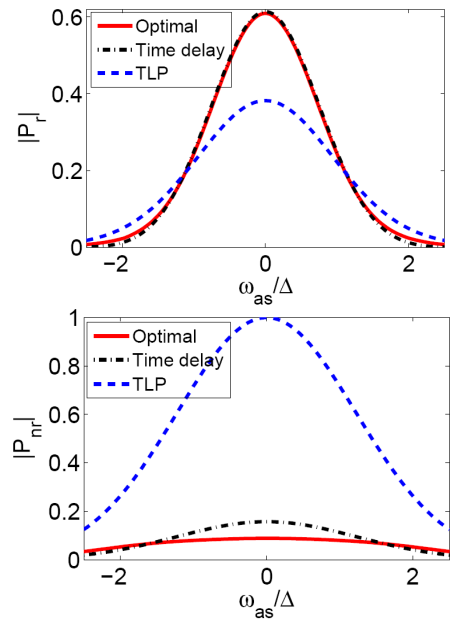}
\par\end{centering}

\caption{\label{fig:Spectrum-of-maximal-Ir-Inr}
The resonant signal and nonresonant background with different phase shaping schemes for the probe pulse, while
keeping pump and Stokes pulses unshaped. The optimal
shaping scheme for maximal $I_{r}-I_{nr}$ (red solid lines), quasi-optimal time delay scheme (black dash dotted lines) and
TLP scheme (blue dashed lines) are shown together for comparison. Parameters are the same
as in Fig.\ref{fig:Optimal-phase-for-Ir-Inr}.}

\end{figure}

To check exactly how much improvement can be made using the optimal phase scheme, the corresponding
CARS spectra of resonant signal $|P_r|$ and nonresonant background $|P_{nr}|$ with different phase shaping schemes are shown
together in Fig.\eqref{fig:Spectrum-of-maximal-Ir-Inr}. Compared with the TLP phase scheme, both the optimal and time delay schemes
can suppress the background to a low level across a broad frequency
band and enhance the resonant signal around $\omega_{as}=0$.
The spectra of $|P_r|$ with the optimal and time delay phase schemes  are almost the same except the tails away from $\omega_{as}=0$, while the spectra  of $|P_{nr}|$
show that the background is suppressed more with the optimal phase scheme especially around $\omega_{as}=0$.
However, this difference in the suppression of $|P_{nr}|$ is small, which leads again to the
conclusion that the time delay scheme is quasi-optimal.  For
other types of objective functionals which place more importance on
broadband background elimination, this conclusion still holds.
Although the time delay scheme could not yield a narrow-band spectrum,
it has almost the best performance in suppressing the broadband non-resonant
background, which is especially advantageous when the background is
very large. In practice, appropriate combinations
of the time delay and other phase shaping schemes could achieve
enhanced performance for signal-background control in CARS.

\section{Conclusions\label{sec:Conclusions} }

Detailed investigations of coherent control for resonant signal
enhancement and non-resonant background elimination of CARS via phase
shaping schemes lead to the conclusion that the maximal resonant signal and
minimal background at a specific frequency may be achieved by
shaping the probe pulse only, while keeping pump and Stokes pulses
unshaped. The optimal probe phase function in two-pulse CARS
is approximately a superposition of linear and arctangent type phases for the pump, which
enhances the resonant signal more than other schemes.
 As a balance of resonant signal enhancement and non-resonant
background suppression, the optimization of the objective functional
$|P_{r}(\Omega_{as})|^{2}-k|P_{nr}(\Omega_{as})|^{2}$ could simultaneously
generate a CARS signal with large resonant component and small background.
To achieve broadband non-resonant background suppression,
the difference of the integrated signal-background intensity over the
entire spectrum is taken as the objective functional. It is found that
the optimal phase shaping configuration consists of unshaped pump
and Stokes pulses and a quasi-time-delay probe pulse.
Numerical simulations show that the background is suppressed more especially around $\omega_{as}=0$ with the optimal phase scheme
than with the time-delay scheme. But the difference in the suppression of $|P_{nr}|$ is small,
which leads again to the conclusion that the well-known time delay scheme is a good approximation for
optimal resonant signal enhancement and broadband background suppression.
It is expected that performing coherent control of the resonant
signal and background could help improve the
performance of CARS spectroscopy and microscopy, especially when using
femtosecond pulses.

There are still many open questions to answer. In this work,
only the pulse shaping in the time domain is considered. The spatial resolution of CARS microscopy is diffraction
limited, without employing quantum effects. Potma \emph{et al.} apply the concept of focus engineering\cite{potma1,potma2,potma3,potma4} to
enhance the sensitivity of CARS microscopy to "chemical interfaces".
However, coherent control strategies
by pulse-shaping in the spatial domain to improve the spatial resolution
of CARS microscopy have not been given much attention.
Another point is chemical selectivity in CARS microscopy. Experiments
have demonstrated that not only can the CARS signals from different vibrational
modes in one molecule interfere but also from different molecules.
By adjusting the pulse phases, the signal from one special molecule
can be effectively enhanced while that of another molecule is suppressed\cite{Konradi:2006p11028}.
Thus, the simultaneous enhanced chemical selectivity and spatial resolution
of CARS microscopy needs to be explored systematically in the
future by phase shaping in the time and spatial domains.\\

\begin{acknowledgments}
This work is supported by National Natural Science Foundation of China
(No. 61074052 and No. 61072032), Open Foundation of State Key Laboratory
of Precision Spectroscopy and Foundation of President of Hefei Institutes
of Physical Science, Chinese Academy of Sciences. Rabitz acknowledges
support from Chinese Academy of Sciences Visiting Professorship
for Senior International Scientists and the US National Science Foundation.
\end{acknowledgments}
\appendix

\section{Variation of resonant Signal}
In the appendices, the following notations are used
for clarity\begin{align*}
\frac{1}{\sqrt{\omega^{2}+\Gamma^{2}}}\sin\left(\Phi_{pr}(\omega)+\alpha(\omega)\right) & =a_{2}(\omega)\\
\frac{1}{\sqrt{\omega^{2}+\Gamma^{2}}}\cos\left(\Phi_{pr}(\omega)+\alpha(\omega)\right) & =b_{2}(\omega)\\
\left\{ \int_{-\infty}^{\infty}e^{-\frac{3x^{2}}{2\Delta^{2}}}\frac{1}{\sqrt{x^{2}+\Gamma^{2}}}\sin\left(\Phi_{pr}(x)+\alpha(x)\right)dx\right\}  & =A_{2}\\
\left\{ \int_{-\infty}^{\infty}e^{-\frac{3x^{2}}{2\Delta^{2}}}\frac{1}{\sqrt{x^{2}+\Gamma^{2}}}\cos\left(\Phi_{pr}(x)+\alpha(x)\right)dx\right\}  & =B_{2}\end{align*}
and \begin{align}
\sin\left(\Phi_{pr}(\omega)\right) & =a_{1}(\omega)\\
\cos\left(\Phi_{pr}(\omega)\right) & =b_{1}(\omega)\\
\left\{ \int_{-\infty}^{\infty}e^{-\frac{3x^{2}}{2\Delta^{2}}}\sin\left(\Phi_{pr}(x)\right)dx\right\}  & =A_{1}\\
\left\{ \int_{-\infty}^{\infty}e^{-\frac{3x^{2}}{2\Delta^{2}}}\cos\left(\Phi_{pr}(x)\right)dx\right\}  & =B_{1}\end{align}

The variational method is used to analyze all the stationary points
for $|P_{r}|^{2}$ with respect to the phase function $\Phi_{pr}(\omega_{pr})$.
The variational condition $\delta\left|P_{r}\right|^{2}=0$ requires

\begin{equation}
\mathbf{Re}(P_{r}^{*}\delta P_{r})=0\label{eq:Necessary-condition}\end{equation}
From this criterion, we have

\begin{equation}
\begin{split}
\mathbf{Re}[\left\{ \int_{-\infty}^{\infty}dx\frac{e^{-\frac{3x^{2}}{2\Delta^{2}}}}{\sqrt{x^{2}+\Gamma^{2}}}\exp
\left[-i\left(\Phi_{pr}(x)+\alpha(x)\right)\right]\right\}\\
\times i\frac{e^{-\frac{3\omega_{pr}^{2}}{2\Delta^{2}}}}{\sqrt{\omega_{pr}^{2}+\Gamma^{2}}}\exp
\left[i\left(\Phi_{pr}(\omega_{pr})+\alpha(\omega_{pr})\right)\right]]=0
\end{split}\end{equation}
which is equivalent to

\begin{equation}
A_2b_1-B_2a_1=0,\label{eq:full-condition}\end{equation}
so the general extremal phase functions $\Phi_{pr}(\omega_{pr})$
for all the critical points of $|P_{r}|^{2}$ satisfy

\[
\Phi_{pr}(\omega_{pr})=L[\omega_{pr}]\pi+\arctan(\omega_{pr}/\Gamma)+constant,\]
where $L[\omega_{pr}]\in\{0,\ 1\}$ is the integer function of $\omega_{pr}$.
It is easy to find that when $L[\omega_{pr}]=\mathrm{0\ or\ 1}$,
\emph{i.e.} $\Phi_{pr}(\omega_{pr})=\arctan(\omega_{pr}/\Gamma)+constant$,
then $|P_{r}|^{2}$ reaches its global maximal value. When $L[\omega_{pr}]=\mathrm{Heaviside}(\omega_{pr})$,
\emph{i.e.} $\Phi_{pr}(\omega_{pr})=\arctan(\omega_{pr}/\Gamma)+\mathrm{Heaviside}(\omega_{pr})\cdot\pi+constant$,
$|P_{r}|^{2}$ has its minimal value 0. Hence, the $\arctan(\omega_{pr}/\Gamma)$\textit{
}\textit{\emph{phase}} function is the optimal solution for maximizing the
peak resonant signal intensity. For the other cases of $K[\omega_{pr}]$,
the phase functions are either local extremal points or saddle points.
However, numerical optimization did not reveal any local extremal
points, which may indicate that the control
landscape\cite{Rabitz:2004p1778} for the resonant signal intensity
is trap-free in this pure phase shaping strategy. A detailed proof of
the trap-free landscape for CARS signal needs to be investigated.

\section{Variation of the difference between the Signal and Background}

For simplicity in the proof, we set $P_{r0}=P_{r}/C$ and $P_{nr0}=P_{nr}/\chi_{nr}$,
and introduce a parameter $\lambda=k\left(\chi_{nr}/C\right)^{2}$,
then the optimization of the objective functional

\begin{equation}
J=\left|P_{r0}\right|^{2}-\lambda\left|P_{nr0}\right|^{2}\label{eq:J}\end{equation}
is equivalent to optimizing the original functional Eq. \eqref{eq:JOri},
because $\left|P_{r}\right|^{2}-k\left|P_{nr}\right|^{2}=C^{2}(\left|P_{r0}\right|^{2}-\lambda\left|P_{nr0}\right|^{2})$.

Following a similar procedure to optimizing $|P_{r}|^{2}$ in Appendix
A, the optimal phase function for $J$ satisfies

\begin{equation}
\lambda(a_{1}B_{1}-A_{1}b_{1})=a_{2}B_{2}-A_{2}b_{2}\end{equation}
hence

\begin{equation}
\tan\left(\Phi_{pr}(\omega)\right)=\frac{a_{1}(\omega)}{b_{1}(\omega)}=\frac{\lambda(\omega^{2}+\Gamma^{2})A_{1}-A_{2}\omega+B_{2}\Gamma}{\lambda(\omega^{2}+\Gamma^{2})B_{1}-B_{2}\omega-A_{2}\Gamma}\label{eq:tan-pr-pnr-full}\end{equation}

According to the expressions for $|P_{r}^{(3)}(\tilde{\omega}_{as})|^{2}$
and $P_{nr}^{(3)}(\tilde{\omega}_{as})|^{2}$, the value of $J$ will
not change if a trivial phase constant is added to $\Phi_{pr}(\omega_{pr})$.
So the general solution for maximizing $J$ can be expressed as the sum
of one special solution and a trivial phase constant. The numerical
simulation in Fig.\eqref{fig:Optimal-different-k} shows that the
optimal phase is just an odd function of $\omega_{pr}$. So $\Phi_{pr}(\omega_{pr})$
to optimize $J$ just makes $\sin\left(\Phi_{pr}(\omega_{pr})\right)$
an odd function of $\omega_{pr}$ and $\cos\left(\Phi_{pr}(\omega_{pr})\right)$
an even function of $\omega_{pr}$ , thus

\begin{eqnarray}
{\scriptstyle A_{1}= }& {\scriptstyle \int_{-\infty}^{\infty}e^{-\frac{3x^{2}}{2\Delta^{2}}}\sin\left(\Phi_{pr}(x)\right)dx =0}\end{eqnarray}

\begin{eqnarray}
{\scriptstyle B_{1}  = }& {\scriptstyle \int_{-\infty}^{\infty}e^{-\frac{3x^{2}}{2\Delta^{2}}}\cos\left(\Phi_{pr}(x)\right)dx\neq0}\end{eqnarray}

\begin{eqnarray}
{\scriptstyle A_{2}=} & {\scriptstyle \int_{-\infty}^{\infty}e^{-\frac{3x^{2}}{2\Delta^{2}}}\frac{1}{x^{2}+\Gamma^{2}}(x\cdot\sin\left(\Phi_{pr}(x)\right)+\Gamma\cos
\left(\Phi_{pr}(x)\right))dx \neq0}\end{eqnarray}

\begin{eqnarray}
{\scriptstyle B_{2}= }& {\scriptstyle \int_{-\infty}^{\infty}e^{-\frac{3x^{2}}{2\Delta^{2}}}\frac{1}{x^{2}+\Gamma^{2}}(x\cdot\cos\left(\Phi_{pr}(x)\right)-\Gamma\cdot\sin\left(\Phi_{pr}(x)\right))dx =0 }\end{eqnarray}

So we have

\begin{eqnarray}
{\scriptstyle \tan\left(\Phi_{pr}(\omega)\right)=\frac{a_{1}(\omega)}{b_{1}(\omega)}=
\frac{-A_{2}\omega}{\lambda(\omega^{2}+\Gamma^{2})B_{1}-A_{2}\Gamma} = \frac{\omega}{\Gamma-\lambda\frac{B_{1}}{A_{2}}(\omega^{2}+\Gamma^{2})}}\end{eqnarray}

The special solution $\Phi_{pr}(\omega)=\arctan(\frac{\omega}{\Gamma-\lambda\frac{B_{1}}{A_{2}}(\omega^{2}+\Gamma^{2})})$
satisfies the assumption, so the general solution for maximal
$J$ is

\begin{eqnarray*}
\Phi_{pr}(\omega) & = & \arctan(\frac{\omega}{\Gamma-\lambda\frac{B_{1}}{A_{2}}(\omega^{2}+\Gamma^{2})})+constant\end{eqnarray*}

Here we define $\gamma=B_{1}/A_{2}$, which could not be determined
analytically and was determined by iteratively solving the equation

\begin{equation}
{\scriptstyle \frac{B_{1}}{A_{2}}=\frac{\left\{ \int_{-\infty}^{\infty}e^{-\frac{3x^{2}}{2\Delta^{2}}}\cos\left(\arctan\left(\frac{x}{\Gamma-\lambda\gamma(x^{2}+\Gamma^{2})}
\right)\right)dx\right\} }{\left\{ \int_{-\infty}^{\infty}e^{-\frac{3x^{2}}{2\Delta^{2}}}\frac{1}{\sqrt{x^{2}+\Gamma^{2}}}\sin\left(\arctan\left(\frac{x}
{\Gamma-\lambda\gamma(x^{2}+\Gamma^{2})}\right)+\alpha(x)\right)dx\right\} }=\gamma}\label{eq:gamma}\end{equation}
Given parameters $\lambda$ ,$\Delta$ and $\Gamma$, we can always
get the numerical value of $\gamma$.

\bibliographystyle{unsrt}

\end{document}